# Compression Performance of Grayscale-based Image Encryption for Encryption-then-Compression Systems

Warit Sirichotedumrong, Tatsuya Chuman and Hitoshi Kiya
*Tokyo Metropolitan University, Asahigaoka, Hino-shi, Tokyo, 191-0065, Japan*

**Abstract**
This paper considers a new grayscale-based image encryption for Encryption-then-Compression (EtC) systems with JPEG compression. Firstly, generation methods of grayscale-based images are discussed in terms of the selection of color space. In addition, a new JPEG quantization table for the grayscale-based images is proposed to provide a better compression performance. Moreover, the quality of both images uploaded to Social Network Services (SNS) and downloaded from SNS, are discussed and evaluated. In the experiments, encrypted images are compressed using various compression parameters and quantization tables, and uploaded to Twitter and Facebook. The results proved that the selection of color space and the proposed quantization table can improve the compression performances of not only uploaded images but also downloaded ones.

## 1. Introduction

Multimedia systems and the Internet have been rapidly growth. A lot of studies on secure, efficient and flexible communications have been reported [1-3]. For securing multimedia data, full encryption with provable security (like RSA, AES, etc) is the most secure options. However, many multimedia applications have been seeking a trade-off in security to enable other requirements, e.g., low processing demands, retaining bitstream compliance, and signal processing in the encrypted domain. To satisfy those requirements, a lot perceptual image encryption schemes has been proposed [4-8].

In order to apply the image encryption to Social Network Services (SNS) or Cloud Photo Storage Services (CPS), encryption schemes have to be compatible to international compression standards, such as JPEG, and also be available to be recompressed by the providers because almost SNS and CPS providers manipulate every uploaded image [9] as illustrated in Fig. 1. Hence, Encryption-then-Compression (EtC) systems [3, 10, 11] which encrypt an image prior to image compression, have been proposed for such scenarios. Although, most studies on EtC systems utilize a proprietary compression scheme which is incompatible with international standards [3, 12-14]. Image encryptions for EtC systems have been proposed to provide the compatibility with international compression standards and availability of recompression [1519].

According to [19], the grayscale-based image encryption, which is the extension of conventional EtC systems [15-18], has been proposed to enhance the robustness against several attacks such as jigsaw puzzle and brute-force attacks [20, 21], and also avoid the effect of color subsampling carried out by SNS and CPS providers. The grayscale-based image encryption firstly generates the grayscale-based image from a full-color image, then the grayscale-based image is encrypted. Furthermore, since the encryption is performed before compression, generation methods of grayscale-based images can be designed freely. In addition, generation methods of grayscale-based images and a compression performance have not been considered. This paper discusses the compression performance affected by generation methods of grayscale-based images. Moreover, since the grayscale-based images used in EtC systems are not strictly grayscale ones, default quantization tables, such as luminance table and chrominance table, are not especially designed based on grayscale-based images. Hence, this paper proposes a new image-dependent quantization table for grayscale-based images to provide the better compression performance for EtC systems.

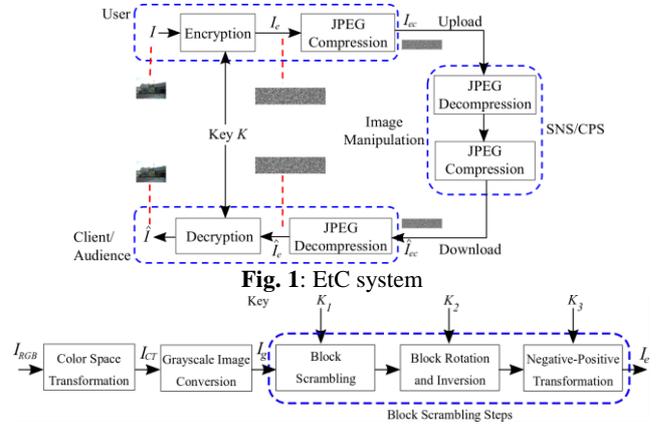

**Fig. 1**: EtC system

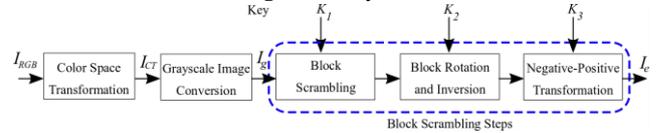

**Fig. 2**: Grayscale-based block scrambling image encryption

Evaluation of the color space selection and the proposed quantization table showed that the color space selection and proposed quantization table enhance the effectiveness of EtC systems in terms of compression performance and image quality

## 2. Preparation

In this section, we summarize the grayscale-based image encryption scheme [19], which has been proposed to avoid the effect of color subsampling in EtC systems with JPEG compression. Besides, the application of the encryption scheme to SNS and CPS providers is described here.

### 2.1. Grayscale-based image encryption

Let us consider a full-color image ($I_{RGB}$) with $M \times N$ pixels. In order to generate the encrypted image ($I_e$), the following processes are performed (See Fig. 2).

**Step1:** $I_{RGB}$ is separated into three color channels: red ($i_r$), green ($i_g$), and blue ($i_b$). Then, $i_r$, $i_g$, and $i_b$ are concatenated horizontally or vertically to generate a grayscale-based image ($I_g$) with $3(M \times N)$ pixels.

**Step2:** $I_g$ with $M_g \times N_g$ pixels is divided into non-overlapping blocks each with $B_x \times B_y$. The number of divided blocks, $N_b$, is expressed by

$$N_b = \left\lfloor \frac{M_g}{B_x} \right\rfloor \times \left\lfloor \frac{N_g}{B_y} \right\rfloor \tag{1}$$

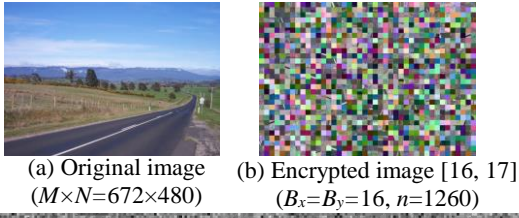

(a) Original image (M×N=672×480)   (b) Encrypted image [16, 17] ($B_x=B_y$=16, n=1260)

(c) Encrypted grayscaled-based image with the proposed scheme from YCbCr color space ($B_x=B_y$=8, n=15120)

**Fig. 3**: Examples of encrypted images

where $\lfloor . \rfloor$ is the round down function to the nearest integer.

**Step3:** Randomly permute the divided blocks based on a random integer which is generated by a secret key $K_1$.

**Step4:** Rotate and invert each divided block randomly based on a random integer generated by a secret key $K_2$.

**Step5:** Perform the negative-positive transformation to each divided block using a random binary integer generated by a secret key $K_3$. A transformed pixel of $i$th block is represented by $p'$ and can be expressed as

$$p' = \begin{cases} p & (r(i)=0) \\ p \oplus (2^L-1) & (r(i)=1) \end{cases} \quad (2)$$

where $r(i)$ is a random binary integer generated by $K_3$ and $p$ is the pixel value of an original image with $L$ bits per pixel.

Note that no color space transformation is carried out in the above steps. Therefore, Fig. 2 meets $I_{RGB}=I_{CT}$. An example of the image encrypted by the conventional encryption scheme [16, 17] is shown in Fig. 3(b).

In JPEG compression, color subsampling is usually done for reducing the color components of a color image. In order to make 8×8-blocks for color subsampling process, the color image must be split into Minimum Coded Unit (MCU) which corresponds to 16×16-blocks. Therefore, the possible smallest block size for the conventional EtC systems is 16×16. On the other hand, grayscale-based images contain only one color channel, so $B_x=B_y$=8 can be utilized as the block size for the encryption scheme. An example of the image encrypted by grayscale-based encryption scheme is shown in Fig. 3(c).

The grayscale-based encryption scheme enables us not only to avoid the effect of color subsampling but also to enhance the security. In terms of the security, the key space of the conventional EtC systems is huge enough against the brute-force attacks [20, 21]. Since the possible smallest block size of grayscale-based block scrambling image encryption is 8×8, and the size of encrypted image is threefold increased, the key space is 12 times larger than that with conventional one.

Considering the jigsaw puzzled solver attacks [20-23], it is difficult to assemble a grayscale-based encrypted image because the number of blocks is larger, the block size is smaller, and there is only one color channel in the encrypted images.

### 2.2. Application to SNS and CPS

It has been confirmed that EtC systems can be applied to SNS [9] as illustrated in Fig. 1. It is also known that almost all SNS providers recompress JPEG images uploaded by users [9, 24, 25]. Therefore, Compression-then-Encryption (CtE) systems cannot be applied to SNS and CPS. On the other hand, EtC systems can be applied to SNS/CPS providers even if the uploaded JPEG images might be recompressed by SNS/CPS providers.

## 3. Compression performance of grayscale-based image encryption

In this section, we propose to improve the compression performance of encrypted images with two new insights: color space transformation and new quantization table for grayscale-based images.

### 3.1. Color space transformation

In the JPEG standard [26], if images have only one color channel, any color space transformation is not performed to be judged as grayscale ones. However, grayscale-based images used in EtC systems are not grayscale ones strictly, so no color space transformation may degrade the compression performance of the images. Because of such a situation, grayscale-based images considering the effects of the color space transformation will be considered in the paper.

In the conventional EtC systems [19], a grayscale-based image $I_g$ is directly produced from the three color components of a color image, i.e. red ($i_r$), green ($i_g$), and blue ($i_b$). On the other hand, we propose to produce $I_g$ from the luminance component ($i_Y$) and two chrominance components ($i_{Cb}$ and $i_{Cr}$) of a color image, as shown in Fig. 2. It is expected that the grayscale-based images have a better compression performance due to the consideration of the color space transformation, even when no color space transformation is carried out in JPEG compression.

### 3.2. Quantization table for grayscale-based images

JPEG softwares, such as Independent JPEG Group (IJG) software [26], generally utilize two default quantization tables to quantize luminance and chrominance of a full-color image, where $i_Y$ is quantized by the luminance quantization table (Y-table), and the chrominance quantization table (CbCr-table) is utilized to quantize $i_{Cb}$ and $i_{Cr}$. However, users are allowed to use other tables rather than the default ones. The image-dependent quantization table has been proposed to minimize the distortion of the quantization process of each block [27]. However, as $I_g$ is produced from $i_Y$, $i_{Cb}$, and $i_{Cr}$, those tables are not designed for $I_g$.

We propose to design a new image-dependent quantization table for $I_g$ called G-table. In JPEG compression, all pixel values in each block of $I_g$ is mapped from [0,255 to [-128,127] by subtracting 128, then each block is transformed using Discrete Cosine Transform (DCT) to obtain DCT coefficients. Since the DCT coefficients are quantized by a quantization table, this paper employs the DCT coefficients for generating G-table. Let $D_n(i, j)$ be the DCT coefficient of the $n^{th}$ block at the position $(i, j)$ where $1 \le i \le 8$ and $1 \le j \le 8$. Considering every block of $I_g$, the Euclidean distance between the origin $O$ and $D_n(i, j)$ is measured, and the arithmetic mean of the distance is expressed by

$$c(i,j) = \frac{1}{N_b}\sum_{n=1}^{N_b}|D_n(i,j)-O| \quad (3)$$

where $I_g$ consists of $N_b$ blocks.

As a set of grayscale-based images which consists of $R$ images is utilized to determine G-table, we define $c_n(i, j)$ as $c(i, j)$ of the $n^{th}$ image and calculate the average of every $c(i, j)$ from $R$ grayscale-based images. $\bar{c}(i, j)$ is calculated as follow.

$$\bar{c}(i,j) = \frac{1}{R}\sum_{n=1}^{R}c_n(i,j) \quad (4)$$

To obtain G-table, $q(i, j)$ represents the quantization step size at $(i, j)$ and is derived from the ratio between $\bar{c}(1, 1)$ and

**Fig. 4**: G-table for grayscale-based images

$\bar{c}(i,j)$. The step size can be calculated by

$$q(i,j) = \left\lceil \frac{\bar{c}(1,1)}{\bar{c}(i,j)} \right\rceil + \varepsilon \quad (5)$$

where $\varepsilon$ is set to 16 for adjusting the Y-table step size at (1, 1) as for IJG software [26].

## 4. Experiments
### 4.1. Experimental set-up
In order to evaluate the compression performance of grayscale-based image encryption for EtC systems, we employed two datasets as below.
(a) 20 images from MIT dataset (672×480) [23]
(b) 1338 images from Uncompressed Color Image Database (UCID) [28]

All images from dataset (a) were encrypted using the proposed scheme with $B_x=B_y=8$. Then, all encrypted images were compressed with 4:4:4 sampling ratio and specific quality factors, $Q_{f_u} \in [70,100]$, using the JPEG standard from IJG software. Note that the images encrypted by the proposed scheme were compressed using G-table, Y-table, or CbCr-table. In addition, JPEG images with $Q_{f_u} \in [70,100]$, were uploaded to Twitter and Facebook as illustrated in Fig. 1. All uploaded images were downloaded afterwards to decode and measure Peak-Signal-to-Noise Ratio (PSNR), respectively.

### 4.2. Results and discussions
*4.2.1. Quantization table for grayscale-based images*

All images in dataset (b) were compressed using IJG software [26] to obtain compressed grayscale-based images. Note that DCT coefficients are extracted during this JPEG compression. According to the procedures in section 3.2, G-table was designed by using the DCT coefficients whereas $N_b=9216$, $R=1338$, and $\varepsilon=16$. As a result, G-table is shown in Fig. 4.

*4.2.2. Quality of uploaded images*

The quality of uploaded images which are images generated from the original ones by using JPEG compression were evaluated based on Rate-Distortion (R-D) curves, which is the relation between the arithmetic mean PSNR of the images and bits per pixel (*bpp*) of JPEG images. The generated JPEG images were respectively decompressed and decrypted, and then the decrypted images are compared with the original ones in order to obtain the PSNR values. Moreover, *bpp* represents the size of the generated JPEG image.

Compared to encrypted JPEG images, $I_{ec}$, from RGB color space, $I_{ec}$ from YCbCr color space contains less bits in one pixel by sacrificing a little PSNR during color space transformation process as illustrated in Fig. 5.

In YCbCr color space, using G-table provides the best R-D compared to other tables and almost the same as that of the unencrypted images with 4:4:4 subsampling ratio. Moreover, Y-table and CbCr-table offer the reasonable R-D.

In RGB color space, G-table gains more compression performance compared to the use of Y-table. However, R-D of encrypted images from RGB color space is incomparable to those from YCbCr color space.

*4.2.3. Quality of downloaded images*

As shown in Fig. 5, encrypted images generated from YCbCr color space outperforms the encrypted ones in RGB

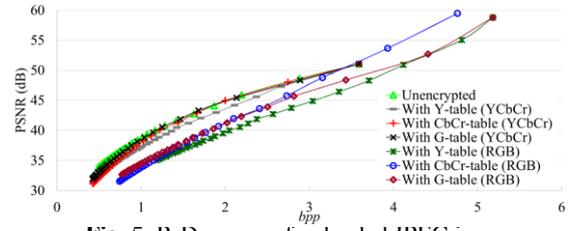

**Fig. 5**: R-D curves of uploaded JPEG images

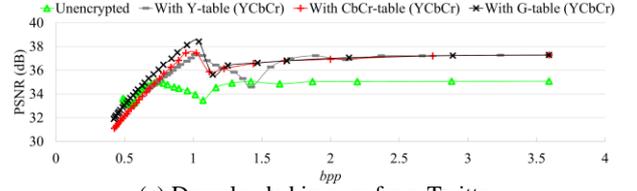

(a) Downloaded images from Twitter

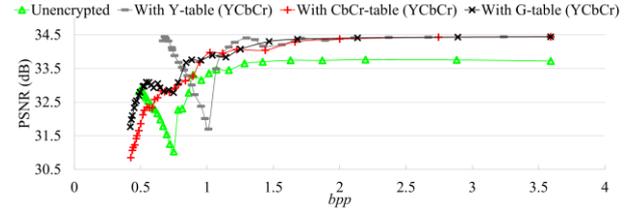

(b) Downloaded images from Facebook

**Fig. 6**: R-D curves of downloaded JPEG images

color space, this section elaborates the application to SNS by comparing the proposed scheme in YCbCr color space with the unencrypted ones. Figure 6 shows R-D curves which represent the relation between PSNR of the downloaded images and *bpp* of the uploaded images.

According to image manipulation carried out by SNS providers [9], the uploaded JPEG images are respectively decoded and compressed based on their specifications when the uploaded images are satisfied the conditions of SNS providers. When $Q_{f_u}$ is high, Twitter recompresses an uploaded JPEG image with 4:4:4 color subsamping ratio to the new JPEG image with $Q_{f_u}=85$ and 4:2:0 ratio. On the other hand, every JPEG image uploaded to Facebook is recompressed to the new JPEG image with a specific quality factor and 4:2:0 ratio.

Since a grayscale-based image contains one color channel, it has the robustness against color subsampling carried out by SNS providers. Meanwhile, other full-color images are distorted by the color-subsampling. As a result, in both providers, the grayscale-based image encryption provides a higher performance than unencrypted images due to the robustness against color subsampling.

Considering the grayscale-based images encryption with various quantization tables, in both providers, when *bpp* is high, the performance of each table is not much different. This is because the JPEG images with high $Q_{f_u}$ are quantized with very low step size, so the images are rarely distorted by the quantization process. Regarding Twitter, when *bpp* is lower, G-table provides the best performance because the uploaded images are not recompressed. Moreover, it is known that most SNS providers utilize Y-table for grayscale images, so using Y-table reduces the quantization error from the recompression carried out by the providers. As a result, some encrypted images uploaded to Facebook, which are quantized by Y-table, offers the better PSNR than other tables as shown in Fig. 6(b).

## 5. Conclusion
This paper introduced the new grayscale-based image encryption for EtC systems with JPEG compression. Firstly,

the generation methods of grayscale-based images were discussed in terms of the color space selection. Moreover, the new image-dependent quantization table for grayscale-based image was proposed to provide a better compression performance. A lot of images was compressed using various compression parameters and quantization tables, and uploaded to Twitter and Facebook in order to evaluate the effectiveness of the EtC systems in terms of image quality and compression performance. The results proved that the selection of color space and the proposed quantization table improve the compression performances of not only uploaded images but also downloaded ones.

**Acknowledgement**

This work was partially supported by Grant-in-Aid for Scientific Research(B), No.17H03267, from the Japan Society for the Promotion Science.